# Spontaneous Valley Splitting and Valley Pseudospin Field Effect Transistor of Monolayer VAgP$_2$Se$_6$

Zhigang Song[1,2,†], Xiaotian Sun[3,†], Jingshan Qi[4], Yangyang Wang[5], Jiaxin Zheng[6], Feng Pan[6], Man-Hong Yung[2], Jinbo Yang[1,7,8*], Jing Lu[1,7*]

[1] State Key Laboratory for Mesoscopic Physics and School of Physics, Peking University, Beijing 100871, P. R. China

[2] Institute for Quantum Science and Engineering and Department of Physics, South University of Science and Technology of China, Shenzhen 518055, China

[3] College of Chemistry and Chemical Engineering, and Henan Key Laboratory of Function-Oriented Porous Materials, Luoyang Normal University, Luoyang 471934, P. R. China.

[4] School of Physics and Electronic Engineering, Jiangsu Normal University, Xuzhou 221116, P. R. China

[5] Nanophotonics and Optoelectronics Research Center, Qian Xuesen Laboratory of Space Technology, China Academy of Space Technology, Beijing 100094, P. R. China

[6] Peking University Shenzhen Graduate School, School of Advanced Materials, Shenzhen University Town, Shenzhen, CN 518055

[7] Collaborative Innovation Center of Quantum Matter, Beijing 100871, P. R. China

[8] Beijing Key Laboratory for Magnetoeletric Materials and Devices，Beijing 100871, P. R. China

## Abstract

Valleytronics is a rising topic to explore the emergent degree of freedom for charge carriers in energy band edges and has attracted a great interest due to many intriguing quantum phenomena and potential application in information processing industry. Creation of permanent valley polarization, i.e. unbalanced occupation at different valleys, is a chief challenge and also urgent question to be solved in valleytronics. Here we predict that the spin-orbit coupling and magnetic ordering allow spontaneous valley Zeeman-type splitting in pristine monolayer of VAgP$_2$Se$_6$ by using first-principles calculations. The Zeeman-type valley splitting can lead to permanent valley polarization after suitable doping. The Zeeman-type valley splitting is similar to the role of spin polarization in spintronics and is a vital requirement for practical devices in valleytronics. The nonequivalent valleys of VAgP$_2$Se$_6$ monolayer can emit or absorb circularly polarized photons with opposite chirality, and thus this material shows a great potential to work as a photonic spin filter and circularly-polarized-light resource. A valley pseudospin field effect transistor (VPFET) is designed based on the monolayer VAgP$_2$Se$_6$ akin to the spin field effect transistors. Beyond common transistors, VPFETs carry information of not only the electrons but also the valley pseudospins.

As a new degree of freedom, valleys have attracted immense attention due to many intriguing quantum phenomena,[1] such as valley magnetism, valley Hall effect,[2-4] and potential applications in information processing industry, since valley polarization was realized in monolayer (ML) $MoS_2$ by circularly polarized optical excitation.[5-7] The principle mechanism of the valley polarization driven by circularly polarized optical excitation is the dynamic creation of non-equilibrium carrier population at two valleys, and thus the lifespan of valley carriers (approximately few picoseconds) is too short.[6-8] Generation of permanent valley polarization thus becomes a central theme and an urgent question to be solved in the valleytronics.[9] Permanent valley polarization may offer an opportunity to realize stable valleytronic devices, such as nonvolatile random access memory based on valleys, and reproduce some phenomena locked by time reversal symmetry, such as valley filtering, circularly polarized electroluminescent and anomalous Hall effect.[10] In analogy to spin Zeeman splitting, lifting valley degeneracy results in Zeeman-type valley splitting, which allows for a permanent valley polarization after a suitable doping.

Recently, several external methods such as applying a magnetic field or electric field,[11] optical Stark effect,[12] magnetic impurities, and magnetic substrates,[13] have been utilized to achieve Zeeman-type valley splitting.[14,15] However, valley degeneracy will be restored and valley polarization will be quenched after external fields or substrates are removed. From a practical perspective, the required field to achieve a sizable valley Zeeman splitting is difficult to realize and unbeneficial to future integrated circuits of valleytronic devices. Magnetic impurities or substrates always have a negative influence on the electron occupation of 2D valleytronic materials due to charge transfer or doping.[16,17] Apparently, it is highly desirable to explore internal (spontaneous) valley polarization (namely, valley ferromagnetism[18]). So far, spontaneous valley polarization is quite rarely reported except a predication on ML $VSe_2$,[19] which suffers from the entangled spins in vicinity of the Fermi level.

In this article, we predict that a narrow-bandgap ferromagnetic semiconductor, monolayer (ML) $VAgP_2Se_6$, has a pair of single-spin massive Dirac cones, which constitute a binary valley degree of freedom in a rectangle cell in contrast to previous honeycomb lattices in valleytronics. The valleys are spontaneously nondegenerate due to spin orbit coupling (SOC) and intrinsic ferromagnetism. Spontaneous valley splitting and pristine stoichiometry promise convenient fabrication of the valleytronic devices. Based on the optical selection rule, ML $VAgP_2Se_6$ can work as a photonic spin filter in the region of terahertz and an ultrathin circularly-polarized-laser, which are long-sought in optics. In analogy to spin field effect transistor, we design a transistor based on this valleytronic material and name it as valley pseudospin field effect transistor (VPFET). The valley current in ML $VAgP_2Se_6$ is tunable under proper bias voltages ($V_{ds}$) and gate voltages ($V_g$) which can carry the chirality characterized by Berry curvature. The chirality can be detected by micro optical Kerr rotation or optical polarization of luminescence. VPFET is a type of multifunctional device beyond common transistors by chirality. The pseudospin field transistors can be treated as a third member of the transistor family besides the electronic transistor and spin transistor. All work here, especially the simulation of the devices, is performed by first-principles calculations in a stable two-dimensional material, and thus within the reach of the current experiments.

The chief novelty of the current result is the following: (1) We find a spontaneous valley Zeeman splitting (or performant valley polarization) in pristine monolayer VAgP2Se6. (2) New-functional devices with the reach of current experiments, especially valley pseudospin transistors, are designed and simulated by the first-principles calculations. Since the valley pseudospin with spontaneous polarization provides a platform for potential kinds of valleytronic devices in analogy to spintronic devices, such as tunneling field effect transistor (TFET) and metal-oxide-semiconductor field effect transistor (MOSFET), our work may open an avenue toward the low-power or high-performance valleytronic devices in the future. (3) We predict a ferromagnetic two-dimensional semiconductor, which is rarely reported before although its corresponding bulk has been synthesized.

**Results and discussions**
**Spontaneous Valley Splitting**

Bulk VAgP$_2$Se$_6$ is stacked with layers consisting of P diatoms, V and Ag atoms centering at the rings of six Se atoms (see Fig. 1).[20] The neighboring P diatoms, V and Ag atoms are bridged through two Se atoms. The equilibrium lattice constants observed by X-ray powder diffraction are $a$ = 6.34 Å, $b$ = 11.18 Å and $c$ = 6.98 Å, respectively. The Interlayers are bonded via weak van der Waals interaction with an interlayer distance of 3.26 Å, allowing layers to be easily exfoliated. Bulk VAgP$_2$Se$_6$ is a ferromagnetic half-metal with a measured Curie temperature of 29 K.[20] Both bulk and ML have the same space group of $C_2$.

The ground state of ML VAgP$_2$Se$_6$ is a ferromagnetic insulator with a magnetic moment of 4 $\mu_B$ per unit cell. According to Heisenberg model, exchange interaction is 63 meV per unite cell. Fig. 2(a) shows the spin-polarized band structure of ML VAgP$_2$Se$_6$ without SOC. The majority-spin (red lines) and minority-spin (blue lines) bands are well split by exchange interaction. The 3$d$ shell of V atom is partially filled, leaving 2 unfilled $d$-orbits in majority spin. The bands in the vicinity of the Fermi level are dominated by majority spin and featured by two massive Dirac cones with a band gap of approximately 39 meV, which constitutes a binary degree of freedom in analogy to the electron spin. The two valleys are degenerate without including SOC. After SOC is included, the corresponding dispersion relation is shown in Fig.

2(b), and the valleys are nondegenerate. SOC decreases the gap of one valley to 29 meV, forming the fundamental bandgap, and increases the gap of the other valley to 44 meV. The two valleys center at the points on the line of $N'-\Gamma-N$, and we label them as $K$ (0, 0.267π) and $K'$ (0, -0.267π), respectively (see Fig. 2(c)). The global bandgap still remains a large value of 29 meV, which is a complex effect of crystal symmetry breaking and SOC. Using the approximation of atomically orbital, SOC can be written as $\lambda L \cdot S$. Orbital momentum is dominated by the component out the material plane, and the SOC is simplified as $\lambda L_z S_z$. In two opposite valleys, the orbital momentum, especially the component vertical to the material plane, has opposite signs according to Figs. 2(d, e), and the spin has the same signs according to the discussions above. Thus the local gaps oppositely responses to the SOC, leading to valley Zeeman-type splitting. The component of in-plane SOC only contributes to a shift in two-dimensional momentum space.

For valley contrast physics discussed below, Berry curvature ($\Omega_n(\boldsymbol{k})$) is one of the key quantities. The integrated Berry curvature in a certain zone ($S$) is proportional to the Hall conductance

$$\sigma_{xy} = \frac{e^2}{2\pi h} \sum_n \int_D \Omega_n(\boldsymbol{k}) d^2\boldsymbol{k}, \tag{1}$$

where the sum is over all occupied states. According to Fig.3, the calculated Berry curvature is valley specific. Berry curvature has an influence on the carriers akin to Lorentz force in the presence of in-plane electric field, resulting in a spatial separation of the carriers at the opposite valleys. This is the valley Hall effect, and the valley Hall effect can be accompanied by transverse valley current without electron current. Berry curvature can thus be directly observed in bulk through subtle signatures of the valley Hall effect. For the valleytronic materials that valleys are well separated, the valley index ($\tau = \pm$) is a good quantum number akin to spin index. The valley Hall conductivity is defined as

$$\sigma_{xy}^v = \sum_\tau \sum_n \frac{e^2}{2\pi h} \int_\tau \tau\Omega_n(\boldsymbol{k}) d^2\boldsymbol{k}, \tag{2}$$

where the sum is over all the occupied states. It characterizes the transverse valley current in presence of an in-plane electric field. The Berry curvature associated with the Kohn Sham Bloch waves is displayed in Fig. 3(b). There is a Berry's flux spreading across each valley and thus a circumnavigation encircling a valley accumulates nonzero Berry phase, leading to nonvanishing valley Hall conductance. If suitably doped, the integration zone can cover only one valley, and the valley Hall effect is equivalent to the anomalous Hall effect. An anomalous Hall effect can be observed by tuning the Fermi level.

**Device Design**

Generally, the different valleys response to photons with opposite polarization. The rectangle color map in the Fig. 3(c) shows the $k$-resolved degree of circular polarization distributed in the Brillion zone, which is defined as[4]

$$\eta(k) = \frac{|p_+(k)|^2 - |p_-(k)|^2}{|p_+(k)|^2 + |p_+(k)|^2} , \qquad (3)$$

where $P_\pm = P_x \pm P_y$, $P_\alpha$ is the canonical momentum transition matrix element connecting the valence and conduction bands. The circular polarization is selective at different valleys, and $\eta(k) \approx \pm 100\%$ for the $K$ and $K'$ valleys, respectively. This implies that a left-hand (or right-hand) polarized photon is selectively absorbed in the $K$ ($K'$) valley, while the charily opposite photon is almost prohibited.

If a beam of the linearly polarized or natural light with photon energy in the energy window between the bandgaps of two valleys is irradiated on ML VAgP$_2$Se$_6$, only photons with certain spin (chirality) are absorbed, and the photons with the opposite spin propagate without any barriers (see Fig. 4(a)). When the magnetization is reversed, the charities of the absorbed and transmitted photons are reversed at the same time. Due to nonequivalent absorption of right- and left-handed photons, this material can work as a photonic spin filter in the certain energy interval. Although the absorption rate of conventional materials is always low, leading to a low polarization of photonic spin. The absorption rate of 2D materials increases with the number of layers. The spin polarization of photons thus grows with the number of layers.

Advanced nanotechnology enables the fabrication of the transistors using ML VAgP$_2$Se$_6$. At first, a device of MOSFET (illustrated in Fig. 4(b)) can be designed, which is easy to realize in current experiments.[21,22] We refer to it as valley pseudospin field effect transistor (VPFET). A double gate herein is utilized. The gates are realized by a metal layer and a dielectric region of Al$_2$O$_3$ with an equivalent oxide thickness of 0.2 nm. The transports are parallel to the C$_2$ axis (*b*-direction shown in Fig. 1(a)) of the ML VAgP$_2$Se$_6$. The transports perpendicular to the C$_2$ axis (*a*-axis in Fig. 1(a)) is trivial for valleys, since the valleys merge in this direction. The atomically thin VAgP$_2$Se$_6$ doped with electrons are used as source and drain electrodes. Without loss of generality, we use a channel as long as possible (up to 6.7 nm) considering our computing capacity. In an excellent accord to the band structure, there is a small bandgap of approximately 29 meV in the transmission spectrum without any bias or gate voltages in low temperatures due to the presence of the nonzero gap of the channel material (see Fig. 4(c)). The small discrepancy between the transmission gap and the band gap (29 meV vs 32 meV) is ascribed to quantum limit effect originating from the finite channel length. The transmission coefficients near the valence band maximum (VBM) and the conduction band minimum (CBM) are dominantly contributed from one valley. The $k$-resolved transmission coefficient of one energy point near VBM ($E$ = -6.5 meV) is shown in the Fig. 4(c). From Fig. 4(d), at this energy only one valley can be transmitted and the other is forbidden to some degree. We define a valley polarization of the transmission coefficient at one energy point $\xi(E, V_{ds}, V_g)$ as:

$$\xi(E,V_{ds},V_g) = \frac{\int_{l+} T_+(E,V_{ds},V_g,k_a)dk_a - \int_{l-} T_+(E,V_{ds},V_g,k_a)dk_a}{\int_{l+} T_+(E,V_{ds},V_g,k_a)dk_a + \int_{l-} T_+(E,V_{ds},V_g,k_a)dk_a} \quad (4)$$

where $T_+$ and $T_-$ are transmission coefficients contributed from the $K$ and $K'$ points, respectively at given energy $E$, wave vector $k_a$, bias voltage $V_{ds}$ and gate voltage $V_g$. the integral intervals $l_+$, and $l_-$ are limited to the vicinity of the $K$ and $K'$ points, respectively. The current between the source and drain at is obtained by the modified Landauer−Bűttiker formula:

$$I(V_{ds},V_g) = \frac{2e}{h} \int_{-\infty}^{+\infty} \int_{-\frac{\pi}{a}}^{\frac{\pi}{a}} \{T(E,V_{ds},V_g,k_a)[f_s(E-\mu_s) - f_d(E-\mu_d)]\}dk_a dE \quad (5)$$

where $T(E,V_{ds},V_g)$ is the transmission coefficient at given energy $E$, wave vector $k_a$, bias voltage $V_{ds}$ and gate voltage $V_g$. $f$ is the Fermi-Dirac distribution function, $\mu_s$ and $\mu_d$ the chemical potential of source and drain, respectively. The temperature of 1 K is used in the source and drain. The valley polarization of the current $\zeta(I, V_{ds}, V_g)$ can be calculated by

$$\zeta(V_{ds},V_g) = \frac{2e}{h} \sum_\tau \int_{-\infty}^{+\infty} \int_l \{\tau T_\tau(E,V_{ds},V_g,k_a)[f_s(E-\mu_s) - f_d(E-\mu_d)]\}dk_a dE / I(V_{ds},V_g) , \quad (6)$$

where the integral interval ($l$) is limited to the vicinity of the valleys in that $\tau$ remains a good quantum number. At a small $V_{ds}$, $\zeta(I, V_{ds}, V_g) \approx \xi(E, V_{ds}, V_g)$. The valley polarization $\xi$ of the transmission coefficient is about 51% at $E = -6.5$ meV. If a long channel is utilized, the quantum limit effect is vanishingly small, and the valley polarization of the transmission coefficient is expected to be larger, even up to $\xi = 100\%$ in vicinity of the Fermi level.

The principles of the VPFETs is similar to the common transistors in electronics. A bias voltage $V_{ds}$ lifts the chemical potentials of left and right electrodes upward and downward $V_{ds}/2$, resulting in a bias window. The gate voltage $V_g$ always results in a lift of the chemical potential in the part of the gate zone. The chemical potential still resides in the bandgap under a small bias voltage of $V_{ds}$ and a gate voltage $V_g = 0$. The valley transmission wave functions in real space are constrained near the source (drain) and hard to reach the drain (source). This leads to a very low current and sets the device in off-state. Fig. 5 displays the transmission characteristics under a small bias voltage of $V_{ds} = 0.02$ V. The transmitted current is $1.97 \times 10^{-4}$ μA/μm under $V_{ds} = 0.02$ V and $V_g = 0$ V. The current is valley-polarized $\zeta \approx 74\%$ from the valley polarized transmission efficient of $\xi = 74\%$ at $E = -10$ meV, as shown in Fig. 5(b). If a suitable gate voltage is applied, the bias window covers the CBM or VBM of either of the valleys and excludes the CBM and VBM of the other valley. The valley transmission wave functions spread over whole channel in real space. Electrons (holes) undergo right (left) acceleration and transmit, resulting in a significantly enhanced valley current and setting the device in an on-state. An on-state current of 1.19μA/μm is obtained under a gate voltage of -0.5 V, corresponding to a large current on/off ratio of $6.31 \times 10^3$. The current has a nearly completely valley polarization in low energy limit in terms of a valley polarization of the transmission coefficient of $\xi = 100\%$ at $E = -10$ meV, as shown in Fig. 5(d). Although the difference between the on-state and off-state valley current could be as large as several orders of magnitude, improper doping approach, channel lengthens, gate voltages and bias voltages will degrade valley polarization of the on-state current. It is interesting that the transmission of the valley current exhibits the dependency on the magnetization directions.

If the left and right electrodes are doped with holes and electrons, respectively, the device becomes a valley pseudospin tunneling field effect transistor (VPTFET), which is shown Fig. 6(a). Carriers on the band extremums can transmit in the channel, and circularly polarized light can be emitted in the channel after inter-band recombination. By careful fabrication of a cascade structure, the quantum cascade radiation can be realized, resulting in a circularly polarized cascade laser.[23] As far as we known, circularly-polarized-light resource has rarely been reported in two-dimensional crystals. When the magnetization is reversed, the other valley is mainly transmitted, and the valley current is reversed. If the spins of the source and drain electrodes orientate opposites, valley pseudospins are opposite at source and drain electrodes, and the valley current is vanishingly small (see Fig. 6(b)). When the spins of the source and drain electrodes are parallel, valley pseudospins are same, and the valley current can be large (see Fig. 6(c)). Thus a valley pseudospin valve can be designed by controlling the magnetization of the electrodes.

For the common field effect transistors (FET), the current is controlled by the gate voltages and bias voltages, while the current coupled to the valleys is controlled by the bias and gate voltages in our valleytronic logical device. Although they seem the same in device architecture, our transistors are distinguished from previous FETs by an advantage of chirality, which is characterized by Berry curvature and optical helicity. The transmitted valleys can be detected by their charities using circularly polarized luminescence or magneto-optical Kerr rotation.[24,25]

Generally, the calculated band structures involved transition metal atoms are dependent of the on-site Coulomb repulsion (Hubbard $U$). Here the Hubbard U is self-consistently determined to 0 using quantum espresso packages.[26,27]. Therefore, our calculated results associated with the valley Zeeman-type splitting and valleytronics devices are reasonable. To further confirm our calculations, we check our band structure with the method of HSE06 (see Fig.2 (f)). The bandgap increases to a value of 1.65 eV, and our main results are valid, especially valley Zeeman-ty valley splitting. Similar results have been reported in previous work.[28,29] The bandgap difference is 35meV at two valleys.

**Methods**

Calculations of electrostatic and optical properties here are performed using Vienna ab initio simulation package in the framework of the projector augmented waves.[30,31] The cutoff energy is set as 520 eV, and a vacuum space larger than 20 Å is applied to ensure decoupling periodic imagines. The generalized gradient approximation (GGA) of Perdew-Burke-Ernzerh[32] form is adopted to describe the exchange and correlation interaction. Quasi-Newton algorithm is used to relax the atoms when the crystal structure is close to the local minima of free energy. All the atoms are relaxed until the force on every atom is smaller than 0.01 eV/ Å using a Γ-centered $9 \times 7 \times 1$ ***k***-point mesh .

All the converged and rigorous predictions of the valley transports are finished in a parameter-free first-principles method, which is different from the ideal concept of devices in previous studies. The device is simulated based on a model of gated two-probe FET using nonequilibrium Green's function approach coupled with density functional theory, which is implemented in ATK 2016 package.[33-35] In the code, the gates and dielectric region are included during self-consistently solving the Possion's and Kohn–Sham's equations by introducing the boundary conditions. Atomically orbital basis sets and SG15 pseudo potential are used, and the SOC is included. A Γ-centered $21 \times 1 \times 75$ ***k***-point mesh is implemented for both the central region and electrodes. Doped ML VAgP$_2$Se$_6$ is used as electrodes.

**Summary**

In summary, the valleys are spontaneously nondegenerate due to the ferromagnetic order and SOC. The Zeenm-type valley splitting is a long-sought property due to its vital importance to valleytronic devices. Considering the optical selection rule and Zeeman-type valley splitting, a long-sought photonic spin filter and circularly-polarized electro laser are designed. Based on the valley splitting, we design a VPFET, which is another member of transistor family. All work here, especially the simulation of devices, is finished in a method of first-principles calculations in a stable monolayer semiconductor and thus possible to be realized by experimental technology in the near further. Our work may open an avenue towards valleytronic devices in analogy to spintronic devices by the application of spintronic concepts in valley pseudospin.

**Author contributions:** Z.G.S., X.T.S, M.H.Y., J.B.Y, J. L. conceived the calculation. Z.G.S. performed the coding and calculations. Z.G.S., X.T.S, M.H.Y., J.B.Y, Jing Lu wrote the manuscript. All authors took part in the discussions and editing of the manuscript.

**Competing financial interests:**

†The authors declare no competing financial interests.

**Corresponding authors:**

Correspondence and requests for materials should be addressed to Man-Hong Yung, Jinbo Yang, or Jing Lu. (email: yung@sustc.edu.cn, jbyang@pku.edu.cn or Jinglu@pku.edu.cn).

**Acknowledgement:**

This work was supported by the National Key Research and Development Program of China (Nos. 2017YFA206303, 2016YFB0700901, 2017YFA0403701, 2013CB93260), the Natural Science Foundation of China (Projects No. 11674005, 11204110, 51371009, 51171001 and 11674132), and National Materials Genome Project (2016YFB0700600).

**Figure captions**

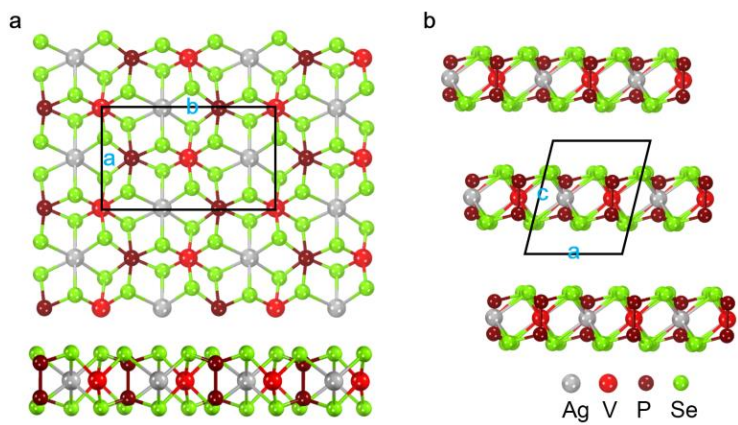

**Figure 1. Crystal structure of VAgP$_2$Se$_6$.** (a) Top (upper) and side (lower) views of monolayer VAgP$_2$Se$_6$. (b) Side view of bulk VAgP$_2$Se$_6$ by layer stacking. The outlines of the monolayer and bulk unit cells are sketched by black lines.

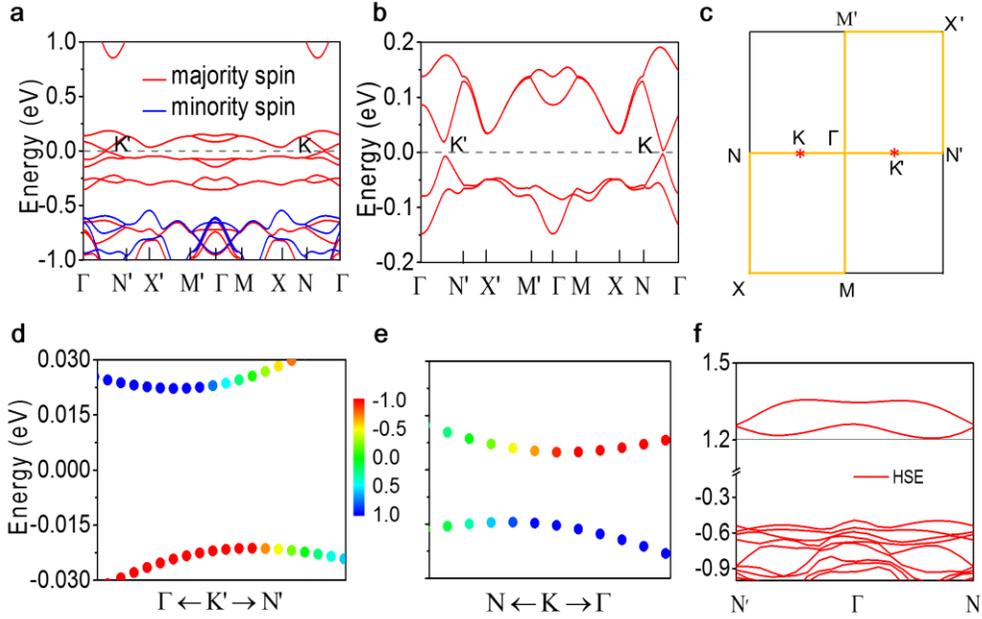

**Figure 2. Electronic structure of monolayer VAgP$_2$Se$_6$.** (a) Spin polarized band structure without including SOC or strain. The band structure in vicinity of Fermi level is fully spin polarized. (b) Band structure near the Fermi level including SOC without strain. (c) Rectangle Brillouin zone and high-symmetry path colored by orange. (d-e) Zoom out of the two valleys colored by the component out plane of orbital momentum. (f) Band structure calculated by HSE06. The Fermi level is set to zero.

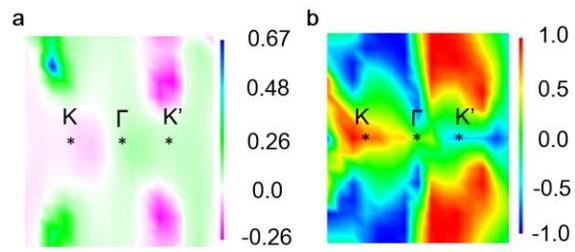

**Figure 3. Valley selectivity of monolayer VAgP$_2$Se$_6$ in the Brillouin zone.** (a) Momentum resolved Kohn-Sham Berry curvature of the whole occupied wave function. (b) Degree of optical circular polarization.

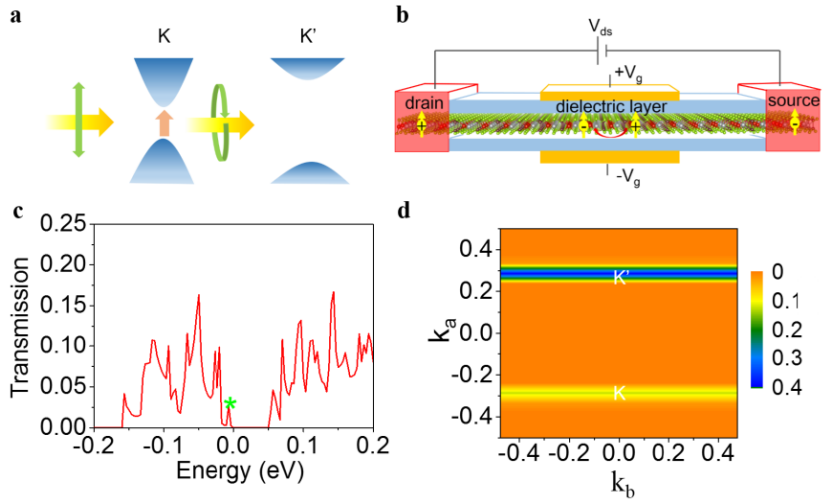

**Figure 4.** Devices of monolayer VAgP$_2$Se$_6$. (a) Schematic of photonic spin filter. (b) Schematic of valley pseudospin metal-oxide-semiconductor field effect transistor. Arrows on the balls index the valley pseudospin. (c) Zero-bias and zero-gate-voltage transmission spectrum (spin is fully polarized) in monolayer VAgP$_2$Se$_6$ with including SOC. (d) ***k***-resolved transmission coefficient of the state with an energy of -6.5 meV, which is labeled by the green star in the transmission spectrum.

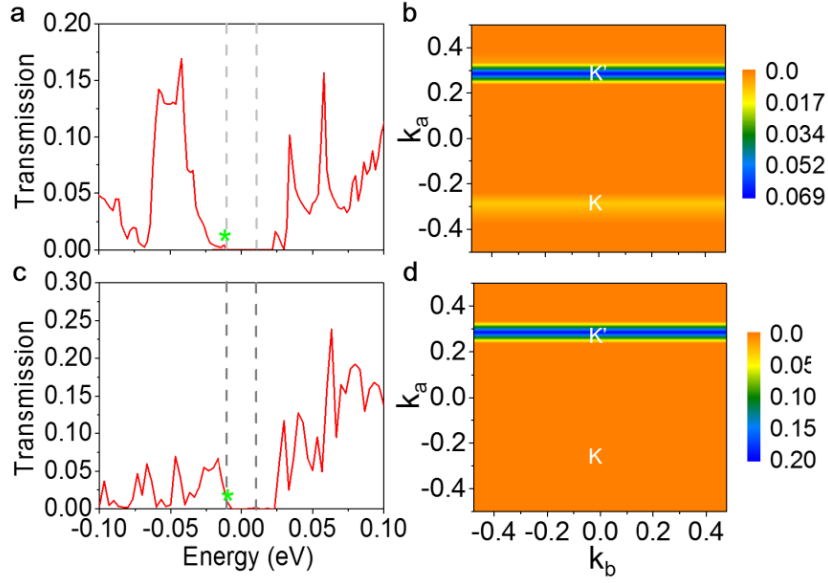

**Figure 5.** Transports of valleytronics devices based on monolayer VAgP$_2$Se$_6$ (a) Transmission spectrum of valley pseudospin field effect transistor under $V_{ds}$ = 0.02 V and $V_g$ = 0 V. (b) *k*-resolved transmission coefficient of the state with the energy of -10 meV, which is indexed by the green star in (b). The plotted transmission state corresponds to the point indexed by the green star in (a). (c) Transmission spectrum under $V_{ds}$ = 0.02 V and $V_g$ = -0.5 V. (d) *k*-resolved transmission coefficient of the state with the energy of $E$ = -10 meV, which is indexed by the green star in (c). SOC is included here. The dash lines represent the chemical potentials of the left and right electrodes. The transmission spectra in the vicinity of the Fermi level are fully spin polarized.

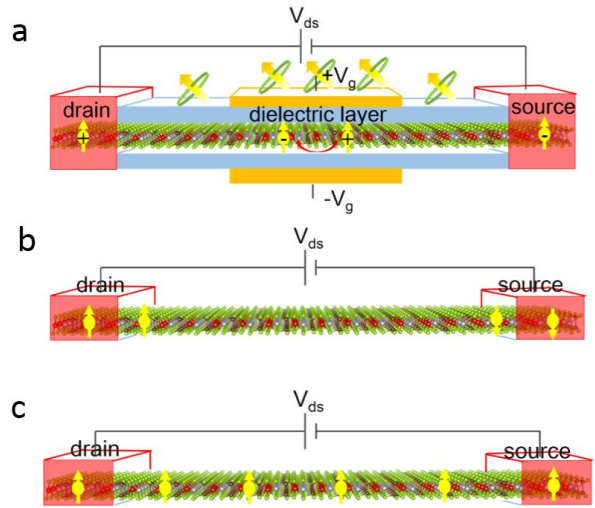

**Figure 6.** (a) Schematic of valley pseudospin tunneling field effect transistor. The plus and minus signs represent holes and electrons, respectively. Arrows in the circles represent the circularly-polarized radiation, and the double arrow illustrates the combination of holes and electrons. (b, c) Schematic of valley pseudospin valve with antiparallel and parallel valley pseudospin. Arrows on the balls index the possible configurations of valley pseudospin.